# Effect of growth conditions on optical properties of CdSe/ZnSe single quantum dots

T. Makino[a*], R. André[a], J.-M. Gérard[b], R. Romestain[a], Le Si Dang[a], M. Bartels[c], K. Lischka[c], and D. Schikora[c]

[a]*Laboratoire de Spectrométrie Physique, CNRS UMR 5588, Université J. Fourier---Grenoble 1, 38402 Saint-Martin-d'Hères Cedex, France*

[b]*CEA-Grenoble, Département de Recherche Fondamentale sur la Matière Condensée/SP2M, 17 avenue des Martyrs, 38054 Grenoble Cedex 9, France*

[c]*Department of Physics, University of Paderborn, Warburger Str. 100, D-33098, Paderborn, Germany*

**Abstract**

**In this work, we have investigated the optical properties of two samples of CdSe quantum dots by using submicro-photoluminescence spectroscopy. The effect of vicinal-surface GaAs substrates on their properties has been also assessed. The thinner sample, grown on a substrate with vicinal surface, includes only dots with a diameter of less than 10 nm (type A islands). Islands of an average diameter of about 16 nm (type B islands) that are related to a phase transition via a Stranski-Krastanow growth process are also distributed in the thicker sample grown on an oriented substrate. We have studied the evolution of lineshapes of PL spectra for these two samples by improving spatial resolution that was achieved using nanoapertures or mesa structures. It was found that the use of a substrate with the vicinal surface leads to the suppression of excitonic PL emitted from a wetting layer.**

Recently, there has been a strong interest for excitons confined in semiconductor quantum dots (QDs). Studies on the temperature dependence of their optical properties are of importance for potential application because the desired device should emit in the visible spectral range at room temperature.

First, we describe photoluminescence (PL) thermo-quenching of CdSe/ZnSe QDs with two different thicknesses (2 ML and 3 ML). These samples were grown using MBE. The GaAs substrate of the thinner sample has a vicinal tilt of $2^o$ to the [111] direction. The thinner sample includes type-A islands that are formed most probably due to strain driven surface kinetics near the island edges. The thicker sample grown on the flat substrate also includes not only type-A but also type-B islands. The conventional PL studies revealed that the thermo-stability up to 150 K was obtained in the thicker sample, whereas rapid temperature-induced quenching starting from the temperature of 6 K was observed in the thinner one. The PL peak energy of the thicker one is significantly higher than that of the thinner sample. Therefore, this broad band results not only from the QDs having inhomogeneous distributions in size and shape but also from the alloyed quantum well (QW)-like layer (otherwise known as a wetting layer). As reported previously, Zn concentration of this QW layer is higher than that of QDs [1]. The relatively rapid thermo-quenching seen in the thinner sample may be induced by the efficient formation of nonradiative centers because, e.g., the ZnSe buffer layer was not obtained in real 2D growth mode.

We have studied the luminescence properties of the 2-ML-thick QD at 6.6 K. These samples were excited by UV lines of an Ar laser. For a 20-micron aperture, averaging over a large area of the sample results in an inhomogeneously broadened spectrum with a full width at half maximum of 65--73 meV, depending on the apertures under examination. In Fig. 1 a micro-PL spectrum is plotted for an aperture size of 500 nm. Reducing the aperture to this size yields a completely different lineshape: The PL spectrum separates in several narrow peaks ($X_1$ to $X_5$) accompanied by low-energy acoustic-phonon sidebands. With an increase in temperature, the sharp peaks broaden, and the relative intensities of the sidebands increase.

We also measured the high-resolution submicro-PL spectrum in the other QD (3 ML grown on the flat substrate) for comparison. These two techniques (a nanoaperture and a mesa) resulted in a same effect. In Fig. 2(a), a submicro-PL spectrum taken at 6.6 K is plotted. Even when reducing the mesa to this size, we observe a broad PL band which acts as a background of several sharp peaks. Figure 2(b) is an expanded view of the low energy part of this spectrum. Because these spikes are superimposed only on the lower energy side of the broad background, the higher-lying PL band may be due to radiative recombination of localized excitons distributed in the QW-like layer. It should be noted that the photon energy region where the sharp peaks are located is similar to that in the thinner sample.

As can be understood from Figs. 1 and 2, the main difference in both the spectra obtained with comparable spatial resolution is the relative PL intensity of the background coming from the wetting layers. It is now well known that the background formation due to the emission from the wetting layer disturbs the experimental observation of photon antibunching from a single QD. Moreover, number of the sharp peaks observed in the thinner QDs is significantly smaller than that in the thicker QDs. We could demonstrate here that, by using a vicinal substrate, much lower spatial resolution (500 nm) than those adopted in previous studies [2] is sufficient to achieve the regime of the single QD spectroscopy. In other words, our experiment resolves the ultranarrow emission from the QDs with estimates of the dot density of 20 dots/micron$^2$, which is one order of magnitude smaller than the literature values.

References
[1] D. Schikola *et al.*, Appl. Phys. Lett. 76, (2000), 418
[2] T. Kuemmel *et al.*, Appl. Phys. Lett. 73, (1998), 3105

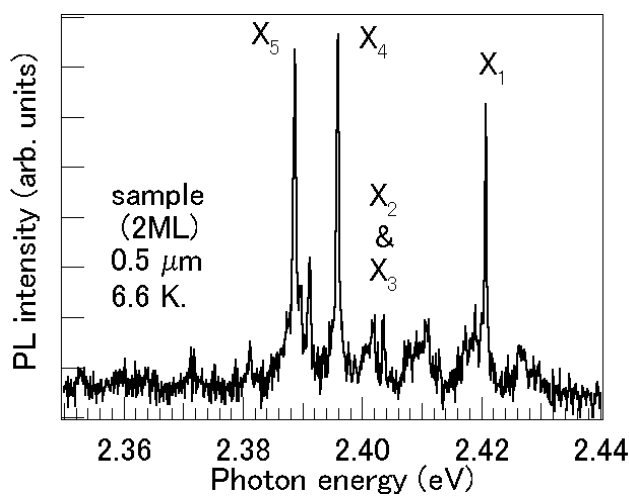

Fig. 1, Submicro-PL spectrum of nano-aperture structures (500 nm) for a nominal CdSe thickness of 2 ML grown on a GaAs substrate with vicinal surface, which has been taken at 6.6K

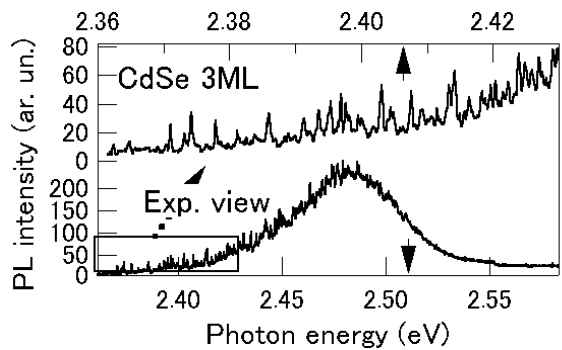

Fig. 2. (a) Submicro-PL spectrum in CdSe nanostructures with thickness of 3 ML grown on a flat substrate. Measured temperature is 6.6 K. (b) Expanded view of the low energy part of the lower trace